\documentclass[twoside,12pt]{article}
\usepackage{indentfirst}
\usepackage{bm}
\usepackage{epsfig}
\usepackage{braket}
\usepackage{float}
\usepackage{amsmath}
\usepackage{graphicx}
\usepackage{amssymb}
\usepackage[dvipdfmx,bookmarks=true,colorlinks,  citecolor=blue,linkcolor=blue,anchorcolor=blue,filecolor=blue,urlcolor=blue,%
           ]{hyperref}
\usepackage{cite}

\begin{document}


\thispagestyle{empty} \vspace*{0.8cm}\hbox
to\textwidth{\vbox{\hfill\huge\sf Commun. Theor. Phys.\hfill}}
\par\noindent\rule[3mm]{\textwidth}{0.2pt}\hspace*{-\textwidth}\noindent
\rule[2.5mm]{\textwidth}{0.2pt}


\begin{center}
\LARGE\bf Angular momentum and parity projected multidimensionally constrained relativistic Hartree-Bogoliubov model
\end{center}

\footnotetext{\hspace*{-.45cm}\footnotesize $^\dag$bnlv@gscaep.ac.cn}

\begin{center}
\rm Kun Wang $^{\rm a,b)}$, \ \ Bing-Nan Lu $^{\rm c)\dagger}$
\end{center}

\begin{center}
\begin{footnotesize} \sl
${}^{\rm a)}$ CAS Key Laboratory of Theoretical Physics, Institute of Theoretical Physics, Chinese Academy of Sciences, Beijing 100190, China\\
${}^{\rm b)}$ School of Physical Sciences, University of Chinese Academy of Sciences, Beijing 100049, China\\
${}^{\rm c)}$ Graduate School of China Academy of Engineering Physics, Beijing 100193, China\\
\end{footnotesize}
\end{center}



\vspace*{2mm}
%
%
%

\begin{abstract}
The nuclear deformations are of fundamental importance in nuclear physics.
Recently we developed a multi-dimensionally constrained relativistic Hartree-Bogoliubov (MDCRHB) model, in which all multipole deformations respecting the $V_4$ symmetry can be considered self-consistently.
In this work we extend this model by incorporating the angular momentum projection (AMP) and parity projection (PP) to restore the rotational and parity symmetries broken in the mean-field level.
This projected-MDCRHB (p-MDCRHB) model enables us to connect certain nuclear spectra to exotic intrinsic shapes such as triangle or tetrahedron.
We present the details of the method and an exemplary calculation for $^{12}$C.
We develop a triangular moment constraint to generate the triangular configurations consisting of three $\alpha$ clusters arranged as an equilateral triangle.
The resulting $^{12}$C spectra are consistent with that from a triangular rigid rotor for large separations between the $\alpha$ clusters.
We also calculate the $B(E2)$ and $B(E3)$ values for low-lying states and find good agreement with the experiments.
%
%
\end{abstract}


\section{Introduction~\label{sec:Introduction}}

Most atomic nuclei are deformed in their ground states~\cite{Bohr1969_1-Nucl.Sruc.,Bohr1975_2-Nucl.Sruc.}.
The nuclear shapes can be described by the surface multipole expansion
\begin{equation}
  R(\theta, \varphi)
  =
  R_0 \left[ 1 + \beta_{00} + \sum_{\lambda=2}^{\infty} \sum_{\mu=-\lambda}^{\lambda}
  \beta_{\lambda\mu}^* Y_{\lambda\mu} (\theta, \varphi) \right],
\end{equation}
where $R$ is the distance
 from the center of mass to the nuclear surface,
 $\lambda$ and $\mu$ are integers characterizing different kinds of multipole deformations.
For example, $\beta_{\lambda0}$'s describe axially symmetric deformations,
the triaxial deformations correspond to $\beta_{\lambda\mu}$ with even $\lambda$'s,
and the tetrahedral shape means $\beta_{32}\neq0$ while all other $\beta_{\lambda \mu}$'s vanish.
These shape parameters can be viewed as generalized coordinates, from which we can build dynamical theories describing the rotations and vibrations of the nuclei.
These collective motions are foundations of the nuclear spectroscopy.


The nuclear spectra are the fingerprints of the underlying nuclear shapes.
The mostly studied deformations are the axially symmetric quadrupole deformation with only $\beta_{20}\neq0$, which manifest itself in the typical spectra of a symmetric rigid rotor, $E \propto I(I+1)$.
A more interesting case is the triaxial deformation with nonzero $\beta_{20}$ and $\beta_{22}$,
which results in the wobbling motion~\cite{Bohr1975_2-Nucl.Sruc.} and the chiral doublet bands~\cite{Frauendorf1997_NPA617-131,Oedegaard2001_PRL86-5866,Starosta2001_PRL86-971}.
The multiple chiral doublet (M$\chi$D) in nuclei with different triaxial configurations were predicted by Meng et al. \cite{Meng2006_PRC73-037303} and recently confirmed experimentally in several nuclei~\cite{Ayangeakaa2013_PRL110-172504,Lieder2014_PRL112-202502,Kuti2014_PRL113-032501,Tonev2014_PRL112-052501,Liu2016_PRL116-112501}. A number of studies have shown that the low-spin signature inversion indicates the existence of triaxiality in certain regions of the nuclide chart~\cite{Bengtsson1984_NPA415-189,Liu1995_PRC52-2514,Liu1996_PRC54-719,Zhou1996_JPGNPP22-415,Riedinger1997_PPNP38-251,Liu1998_PRC58-1849}. 
On the other hand, the parity-breaking octupole deformations $\beta_{3\mu}\neq0$ are also found
in many experiments~\cite{Butler1996_RMP68-349,Moeller2008_ADNDT94-758,Agbemava2016_PRC93-044304}.
For example, the low-lying alternative parity bands found in actinides and some rare-earth nuclei hint the possible existence of reflection asymmetric shapes \cite{Shneidman2003_PRC67-014313,Shneidman2006_PRC74-034316,Wang2005_PRC72-024317,Yang2009_CPL26-082101,Robledo2011_PRC84-054302,Zhu2012_PRC85-014330,Nomura2013_PRC88-021303R,Nomura2015_PRC92-014312}. Direct evidences of a static reflection asymmetric shape were found in $^{224}$Ra \cite{Gaffney2013_N497-199}, $^{144}$Ba \cite{Bucher2016_PRL116-112503} and $^{228}$Th \cite{Chishti2020_NP16-853}.

In recent years some exotic nuclear deformations have attracted a lot of theoretical interests. For example, the reflection asymmetric deformations combined with the triaxial deformations have been proposed to influence the nuclear fission barrier ~\cite{Lu2012_EWC38-05003,Lu2014_PRC89-014323,Lu2014_PS89-054028,Zhao2015_PRC91-014321,Chai2019_CTP71-067}.
The nuclear clustering has been proved to be important in light nuclei, especially in the states near $n\alpha$ binding threshold~\cite{Freer2007_RPP70-2149,vonOertzen2006_PR432-43,Freer2018_RMP90-035004,Feng2018_SCPMA62-12011}.
The nuclear clustering configurations with various arrangement of $\alpha$ clusters are naturally accompanied by exotic shape deformations.
It implies that a proper theoretical description requires simultaneous inclusion of various shape degrees of freedom.
It has been demonstrated that the density functional theories are able to describe the
clustering structure in light nuclei
~\cite{Arumugam2005_PRC71-064308,Maruhn2006_PRC74-044311,Maruhn2010_NPA833-1,Reinhard2011_PRC83-034312,Ichikawa2011_PRL107-112501,Ebran2012_N487-341,Ebran2014_PRC90-054329,Zhao2015_PRL115-022501,Zhou2016_PLB753-227}. The study in Ref.~\cite{Arumugam2005_PRC71-064308} discussed the clustering structure of light nuclei with $N=Z$ by using the axially deformed relativistic mean field model.
By using the cranking covariant density functional theory (CDFT), the rod-shaped carbon isotopes at high spins were studied~\cite{Zhao2015_PRL115-022501,Ren2019_SCPMA62-112062,Ring2019_SCPMA62-112063}.
The studies based on the algebraic model~\cite{Bijker2002_AP298-334} and the antisymmetrized molecular dynamics (AMD) model~\cite{Kanada-Enyo2007_PTP117-655} suggested that the $\alpha$ clustering can appear in the low-lying states of $^{12}$C and other light even-even nuclei~\cite{Marin-Lambarri2014_PRL113-012502,Bijker2014_PRL112-152501,Bijker2021_NPA1006-122077}. In order to study how the nuclei cluster and how the clustering influences the nuclear properties, we need theories that can describe both the excited states and the ground state, and include as many shape degrees of freedom as possible.

The CDFT gives a universal description of the nuclei all over the nuclide chart~\cite{Reinhard1989_RPP52-439,Ring1996_PPNP37-193,Vretenar2005_PR409-101,Meng2006_PPNP57-470,Meng2011_PP31-199,Niksic2011_PPNP66-519,Meng2015_JPG42-093101,Liang2015_PR570-1,Meng2016_RDFNS}.
In CDFT, the Lorentz invariance nature of the Lagrangian automatically determines the spin-orbit interaction, and reveals the origin of the pseudospin symmetry \cite{Ginocchio1997_PRL78-436,Sugawara-Tanabe1998_PRC58-R3065,Meng1998_PRC58-R628,Meng1999_PRC59-154,Lu2012_PRL109-072501,Lu2013_PRC88-024323,Liang2015_PR570-1} in the Fermi sea and the spin symmetry in the Dirac sea \cite{Zhou2003_PRL91-262501,He2006_EPJA28-265}.
An advantage of studying nuclear clustering with CDFT is that it does not presume the existence of $\alpha$ clusters {\it a priori} but generates such configurations from microscopic interactions.
However, the mean field approximation used in CDFT brings the problem that the intrinsic
 symmetries and conservation laws such as the translational invariance, rotational invariance, parity and particle number conservations, are usually broken~\cite{Ring1980_TNMBP}.
One solution is to restore these symmetries with the projection method, examples are the angular momentum projection (AMP) and the parity projection (PP) methods~\cite{Niksic2011_PPNP66-519}.
Combining AMP and PP with the generator coordinate method, we can calculate the rotational-vibrational spectra, which can be directly compared with the experiments~\cite{Niksic2011_PPNP66-519,Yao2010_PRC81-044311,Yao2011_PRC83-014308,Sun2021_PRC104-064319,Sun2021_SB66-2072}. 

In recent years, the multidimensionally constrained covariant density functional (MDCCDFT) theories~\cite{Lu2012_PRC85-011301,Lu2014_PRC89-014323,Zhao2017_PRC95-014320,Zhou2016_PS91-063008
} have been established and applied successfully in several topics.
In these models, the reflection asymmetric and axially asymmetric deformations can be considered
 simulteneously, with only a spatial $V_4$ symmetry remains unbroken for the densities and potentials.
Depending on how to deal with the pairing correlation, two variants was developed: multidimensionally constrained relativistic mean field model (MDCRMF) \cite{Lu2012_PRC85-011301,Lu2014_PRC89-014323} and multidimensionally constrained relativistic Hartree-Bogoliubov model (MDCRHB) \cite{Zhao2017_PRC95-014320}.
The MDCCDFTs have been applied to study the potential energy surfaces and fission barriers in actinides \cite{Lu2012_PRC85-011301,Lu2012_EWC38-05003,Lu2014_PRC89-014323,Lu2014_PS89-054028} and superheavy nuclei \cite{Meng2019_SCPMA63-212011}, the third minima on PESs of light actinides \cite{Zhao2015_PRC91-014321}, the higher order deformations in superheavy nuclei \cite{Wang2021_CPC-in_press}, the non-axial octupole $Y_{32}$ correlations in $N=150$ isotones \cite{Zhao2012_PRC86-057304} and Zr isotopes \cite{Zhao2017_PRC95-014320}, and the axial octupole $Y_{30}$ correlations in M$\chi$D \cite{Liu2016_PRL116-112501}.
The fission dynamics in actinides was studied based on the PESs calculated with the MDCCDFTs \cite{Zhao2015_PRC92-064315,Zhao2016_PRC93-044315,Zhao2019_PRC99-014618,Zhao2019_PRC99-054613,Zhao2020_PRC101-064605}. Furthermore, by including the strangeness degrees of freedom, the MDCCDFTs was employed to study the shape evolution~\cite{Lu2011_PRC84-014328,Lu2014_PRC89-044307} and hyperon pairing correlation~\cite{Rong2020_PLB807-135533} in hypernuclei, and build new nucleon-hyperon effective interactions~\cite{Rong2021_PRC104-054321}.

In this work we focus on the MDCRHB model and its extensions.
Mean field state $\ket{\phi(q)}$ can be obtained by solving the relativistic Hartree Bogoliubov equation in the intrinsic frame. Here $q$ is a collection of quantum numbers specifying the deformation of the state or other intrinsic properties.
Several intrinsic symmetries may be broken due to the mean field approximation
and need to be restored using the symmetry projection methods.
For example, the angular momentum projection for a relativistic mean field model has been implemented
upon a triaxially deformed ground state~\cite{Yao2009_PRC79-044312}
 and an axially deformed reflection-asymmetric ground state~\cite{Yao2015_PRC92-041304R,Xia2018_SCPMA62-42011}.
The generating coordinate method has also been included to
 treat soft potential energy surfaces~\cite{Yao2010_PRC81-044311, Yao2011_PRC83-014308}.
In this work, we incorporate the AMP and PP methods into the MDCRHB model.
Our implementation differs from the previous works in two aspects.
First, we apply a unified framework to treat both the triaxial and octupole deformations.
While in Ref.~\cite{Yao2009_PRC79-044312} the triaxial deformations are considered with an anisotropic 3-dimensional harmonic oscillator (3DHO) basis,
in this work we employ an axially symmetric harmonic oscillator basis in all cases.
This setting allows us to simultaneously break as many symmetries as possible and consider exotic shapes like the triangle or tetrahedron.
Second, we use the Bogoliubov transformation together with a separable pairing force of finite range
to treat the pairing effect.
This makes our results more reliable for extreme deformations for which the pairing force
can play key roles in cluster formations.

With the projected MDCRHB model (p-MDCRHB), we are able to explore various exotic nuclear  shapes.
One direct application is to search for possible $\alpha$-clustering structure in light nuclei.
As the first step toward this direction, in this work we only consider the AMP and PP methods.
The particle number projection method and generating coordinate method also play
essential roles in some circumstances and will be considered in future works.
In Sec.~\ref{sec:Framework}, we present the theoretical framework of the p-MDCRHB model.
In Sec.~\ref{sec:result}, we present the numerical check and the results for the $\alpha$-conjugate nucleus $^{12}$C.
We summarize this work in Sec.~\ref{sec:summary}.

\section{Theoretical Framework}\label{sec:Framework}
The CDFTs solve the nuclear many-body problems in the framework of the density functional theories.
We design a general energy density functional $E[\rho(\bm{r})]$ with $\rho(\bm{r})$ the intrinsic nucleon densities,
then the ground state energy and densities are given by the variational principle $\delta E/\delta \rho = 0$.
In most CDFTs this framework is formally equivalent to a mean field approximation for an
effective Lagrangian,
in which the nucleons are Dirac fermions interacting indirectly via meson exchange  or directly through
zero-range point-coupling interactions~\cite{Nikolaus1992_PRC46-1757,Buervenich2002_PRC65-044308}.
To give correct saturation properties of the symmetric nuclear matter, the nonlinear self-coupling terms \cite{Boguta1977_NPA292-413,Brockmann1992_PRL68-3408,Sugahara1994_NPA579-557} or the density dependence of the coupling constants \cite{Fuchs1995_PRC52-3043,Niksic2002_PRC66-024306} were introduced and applied extensively.
Accordingly, the framework of CDFT can be formulated in four representations: effective interactions with meson exchange (ME) or point-coupling (PC) combined with nonlinear self-coupling (NL) or density dependence of coupling constants (DD).
In MDCCDFT we implemented all four possibilities.
In this work, we illustrate the formalism of the projected MDCRHB model using the density-dependent point-coupling interaction DD-PC1~\cite{Niksic2008_PRC78-034318}.
The formalisms for other parameters are similar and omitted for brevity.

\subsection{Relativistic Hartree-Bogoliubov model}
In this section we briefly introduce the theoretical framework of the MDCRHB model with density-dependent point-coupling interactions.
We start with the effective Lagrangian
\begin{equation}
\begin{split}
  \mathcal{L}
  =&~
  \bar{\psi} ({\rm i} \gamma_\mu \partial^\mu - M) \psi \\
  &
  - \frac{1}{2} \alpha_S (\rho) \rho_S^2
  - \frac{1}{2} \alpha_V (\rho) j_V^2
  - \frac{1}{2} \alpha_{TV} (\rho) \vec{j}_{TV}^2 \\
  &
  - \frac{1}{2} \delta_S (\partial_\nu \rho_S) (\partial^\nu \rho_S) \\
  &
  - e A_\mu j^\mu_C
  - \frac{1}{4} F^{\mu \nu} F_{\mu \nu},
\end{split}
\label{eq:RMF_Lagrangian}
\end{equation}
where $M$ is the nucleon mass, $\alpha_S(\rho)$, $\alpha_V(\rho)$  and $\alpha_{TV}(\rho)$ are density dependent coupling constants in scalar, vector and isospin-vector channels, respectively.
$\rho_S,~j_V$ and $\vec{j}_{TV}$ are the corresponding densities and four-currents. $\delta_S$ is the coupling constant for the derivative term.
$j^\mu_C$ is the proton current, $A_\mu$ is the electromagnetic field and $F_{\mu\nu} = \partial_\mu A_\nu - \partial_\nu A_\mu$ is the field strength tensor.
For DD-PC1 functional~\cite{Niksic2008_PRC78-034318} the functions $\alpha_S(\rho)$, $\alpha_V(\rho)$ and $\alpha_{TV}(\rho)$ share the same ansatz
\begin{equation}
  \alpha_i(\rho) = a_i + (b_i + c_i x)e^{-d_i x} \quad (i = S, V, TV),
\end{equation}
with $x=\rho/\rho_{\rm sat}$, where $\rho_{\rm sat}$ denotes the saturation density of the symmetric nuclear matter.
The constants $a_i$, $b_i$ and $c_i$ are parameters fitted to the experiments.

The pairing correlations are included with the Bogoliubov transformation.
According to the mean field approximation, we take the Bogoliubov vacuum as the trial wave function and solve the variational problem to find the ground state.
In order to prevent the system from collapsing to the negative energy Dirac sea,
we apply the ``no-sea'' approximation, i.e., only the positive energy states are kept in calculating the densities and currents.
The resulting RHB equation~\cite{Ring1996_PPNP37-193,Kucharek1991_ZPA339-23} in coordinate space is
\begin{equation}
  \int {\rm d}^3 \bm{r}'
  \left( \begin{array}{cc}
    h - \lambda & \Delta \\
    -\Delta^* & -h + \lambda
  \end{array} \right)
  \left( \begin{array}{c}
    U_k \\
    V_k
  \end{array} \right)
  =
  E_k
  \left( \begin{array}{c}
    U_k \\
    V_k
  \end{array} \right),
  \label{eq:RHB}
\end{equation}
where $E_k$ is the quasi-particle energy and $\lambda$ is the chemical potential.
$U_k = U_k(\bm{r}^\prime\sigma^\prime)$ and $V_k = V_k(\bm{r}^\prime\sigma^\prime)$ are quasi-particle wave functions.
$\Delta = \Delta(\bm{r}\sigma, \bm{r}^\prime\sigma^\prime)$ is the pairing potential and $h$ is the single particle Hamiltonian
\begin{equation}
  h
  =
  \bm{\alpha} \cdot [ \bm{p} - \bm{V}(\bm{r}) ]
  +
  \beta [ M + S(\bm{r}) ] + V_0 (\bm{r}) + \Sigma_R (\bm{r}),
\end{equation}
where the scalar potential $S(\bm{r})$, vector potential $V^\mu(\bm{r}) = (V_0(\bm{r}), \bm{V}(\bm{r}))$ and the rearrangement term $\Sigma_R(\bm{r})$ are expressed with various nuclear densities and currents,
\begin{subequations}
\begin{align}
  S
  &=
  \alpha_S (\rho) \rho_S + \delta_S \Delta \rho_S, \\
  V^\mu
  &=
  \alpha_V(\rho) j_V^\mu + \alpha_{TV} (\rho) j_{TV}^\mu \cdot \vec{\tau} + e \frac{1-\tau_3}{2} A^\mu, \\
  \Sigma_R
  &=
  \frac{1}{2} \frac{\partial \alpha_S}{\partial \rho} \rho_S^2 + \frac{1}{2} \frac{\partial \alpha_V}{\partial \rho} j_V^2 + \frac{1}{2} \frac{\partial \alpha_{TV}}{\partial \rho} \vec{j}_{TV}^2.
\end{align}
\end{subequations}
The pairing potential $\Delta$ is written as
\begin{equation}\label{eq:pairing_pot}
\begin{split}
  & \Delta_{p_1p_2}(\bm{r}_1\sigma_1, \bm{r}_2\sigma_2) \\
  & = \int {\rm d}^3 \bm{r}'_1 {\rm d}^3 \bm{r}'_2 \sum_{\sigma'_1 \sigma'_2}^{p'_1 p'_2}
  V_{p_1 p_2,p'_1 p'_2}^{\mathrm{pp}} (\bm{r}_1 \sigma_1, \bm{r}_2 \sigma_2, \bm{r}'_1 \sigma'_1, \bm{r}'_2 \sigma'_2)
  \times \kappa_{p'_1 p'_2} (\bm{r}'_1 \sigma'_1, \bm{r}'_2 \sigma'_2)
\end{split}
\end{equation}
with the pairing tensor $\kappa =  \langle \hat{a}(\bm{r}_1\sigma_1) \hat{a}(\bm{r}_2\sigma_2) \rangle $.
In Eq.~(\ref{eq:pairing_pot}) and what follows we use the symbols with and without prime to denote the quantities for initial and final states, respectively.
As a standard procedure in RHB models~\cite{Serra2002_PRC65-064324}, we only include the large component of the Dirac spinors in calculating the pairing tensor.

In this work we use a separable pairing force of finite range~\cite{Tian2009_PLB676-44, Tian2009_PRC79-064301}
\begin{equation}
  V(\bm{r}_1 \sigma_1, \bm{r}_2 \sigma_2,\bm{r}'_1 \sigma'_1, \bm{r}'_2 \sigma'_2)
  = -G \delta(\bm{R}-\bm{R}') P(r) P(r') \frac{1}{2} (1-P_\sigma),
  \label{eq:pairing_force}
\end{equation}
where $G$ is the pairing strength, $(1 - P_\sigma)/2$ is the projector onto the total spin $S=0$,
$\bm{R} = (\bm{r}_1 + \bm{r}_2)/2$ and
$\bm{r} = \bm{r}_1 - \bm{r}_2$
are center of mass coordinate and relative coordinate, respectively.
$P(r)$ is a normalized Gaussian function $P(r) = (4\pi a^2)^{-3/2}\exp\left(-r^2/(4a^2)\right)$
with $a$ an adjustable parameter specifying the range of the pairing force.
The details of calculating the pairing tensor and pairing potential in the ADHO basis can be found in Ref. \cite{Zhao2017_PRC95-014320}.
In this illustrative work we fix the pairing strength to $G=1.1G_0$ with $G_0=728.0$ MeV$\cdot$fm$^{3}$ and the range to $a=0.644$ fm, in which $G_0$ and $a$ are determined by fitting to the pairing gap in the nuclear matter~\cite{Tian2009_PLB676-44,Tian2009_PRC80-024313}.

The RHB equation (\ref{eq:RHB}) can be solved self-consistently starting from a properly chosen initial state.
The resulting minima represent the ground state or the shape isomers.
To obtain the potential energy surfaces $E(\beta)$ we use a modified linear constraint method~\cite{Lu2014_PRC89-014323}.
The Routhian with deformation constraints reads
\begin{equation}
  E'
  =
  E_{\mathrm{MF}} + \sum_{\lambda \mu} \frac{1}{2} C_{\lambda \mu} \beta_{\lambda \mu},
\end{equation}
where $E_{\rm{MF}}$ is the energy functional obtained with the Lagrangian (\ref{eq:RMF_Lagrangian}) and the pairing force (\ref{eq:pairing_force}).
The variable $C_{\lambda \mu}$ varies during the iteration. $\beta_{\lambda \mu}$ is the nuclear deformation
\begin{equation}
  \beta_{\lambda \mu}
  =
  \frac{4 \pi}{3 A R^\lambda} \langle Q_{\lambda \mu} \rangle,
  \label{eq:multipole_constraint}
\end{equation}
where $R = 1.2 A^{1/3}$ fm is the approximate nuclear radius, $A$ is the number of nucleons and $Q_{\lambda \mu} = r^\lambda Y_{\lambda\mu} (\Omega)$, where $Y_{\lambda \mu}$ is the spherical harmonics, is the multipole moment operator.

\subsection{Axially symmetric harmonic oscillator basis}

In the MDCRHB model, we solve the RHB equation (\ref{eq:RHB}) by expanding the spinor wave functions $U_k(\bm{r}\sigma)$ and $V_k(\bm{r}\sigma)$ in an axially deformed harmonic oscillator (ADHO) basis \cite{Ring1997_CPC105-77,Lu2014_PRC89-014323,Zhao2017_PRC95-014320},
\begin{subequations}
  \begin{equation}
    U_{k}(\bm{r} \sigma)
    =
    \left(\begin{array}{c}
    \sum_{\alpha} f_{U}^{k \alpha} \Phi_{\alpha}(\bm{r} \sigma) \\
    \sum_{\alpha} g_{U}^{k \alpha} \Phi_{\alpha}(\bm{r} \sigma),
    \end{array}\right)
  \end{equation}
  \begin{equation}
    V_{k}(\bm{r} \sigma)
    =
    \left(\begin{array}{c}
    \sum_{\alpha} f_{V}^{k \alpha} \Phi_{\alpha}(\bm{r} \sigma) \\
    \sum_{\alpha} g_{V}^{k \alpha} \Phi_{\alpha}(\bm{r} \sigma),
    \end{array}\right)
  \end{equation}
\end{subequations}
where $\Phi_{\alpha}(\bm{r} \sigma)$ is the basis wave function generated by solving the Schrodinger equation with an ADHO potential.
Here we impose the $V_4$ symmetry, which means that the densities and potentials are invariant under the mirror reflections
\begin{equation}
\begin{split}
  \hat{S}_x \phi(x,y,z) &= \phi(-x,y,z), \\
  \hat{S}_y \phi(x,y,z) &= \phi(x,-y,z),
\end{split}
\end{equation}
where $\phi$ represents any scalar field or spatial distribution in the theory.
It immediately follows that the rotation $\hat{S}_z(\pi)=\hat{S}_x\hat{S}_y$ around the $z$-axis by $\pi$ is also a symmetry operation.

In the ADHO basis, the reflection asymmetric shapes can be included by mixing the basis states with different parities.
While the axially deformed shapes preserve the $z$-axis angular momentum projection $K$, the triaxial shapes mix different $K$ blocks and we need to calculate the corresponding mixing matrix elements.
In this case, we expand the densities and potentials in terms of the Fourier series of the azimuthal coordinate $\varphi$ in cylindrical coordinates.
Under the restriction of the $V_4$ symmetry, the expansion for any field $f$ writes
\begin{equation}\label{eq:Fourier_Trans}
  f(\rho, \varphi, z)
  =
  f_0 (\rho, z) \frac{1}{2 \sqrt{\pi}}
  +
  \sum_{n=1}^{\infty} f_n(\rho, z) \frac{1}{\sqrt{\pi}} \cos(2 n \varphi).
\end{equation}
The ADHO basis are also eigenfunctions of the simplex operator $\hat{S}_z(\pi)$ with eigenvalues $S=\pm 1$.
The basis with $S=-1$ can be calculated with a time-reversal operation on those with $S=1$.
For a system with time-reversal symmetry, the single particle energy levels have the Krammer degeneracy and we only need to diagonalize the Hamiltonian in the $S=1$ subspace.

In mean field calculations we need to calculate the potentials on a cylindrical lattice.
The number of mesh points in $z$- and $\rho$- directions are $n_z = 24$ and $n_\rho = 12$, respectively.
In MDCCDFT, the triaxial deformation is included by the Fourier transformation given in Eq. (\ref{eq:Fourier_Trans}).
Choosing $z$ as the principal axis, only the $n=0$ Fourier component survives in the axially symmetric case, which simplifies the calculation regardless of $\varphi$.
A triangular deformed system lying on the $x$-$z$ plane is not axially symmetric about the intrinsic $z$-axis. In such a system, $n>1$ components in the Fourier transformation needs to be taken into account. Due to the $V_4$ symmetry restriction, we only need to calculate the components with even $n$ in the Fourier transformation. The $V_4$ symmetry can simplify the calculation into $1/4$ of the full space with respect to $\varphi$ and we choose the number of points for $\varphi$ integral as $n_\varphi = 12$ for $0 < \varphi < \pi /2$.

\subsection{Angular momentum and parity projections}

The angular momentum projection operator for a general wave function breaking both the axial symmetry and reflection symmetry can be written as
\begin{equation}
  \hat{P}_{MK}^I
  =
  \frac{2I + 1}{8 \pi^2} \int {\rm d} \Omega D_{MK}^{I*} (\Omega) \hat{R} (\Omega),
\end{equation}
where $D_{MK}^I (\Omega) = {\rm e}^{-iM\alpha} d_{MK}^I (\beta) {\rm e}^{-iK\gamma}$ is the Wigner function with Euler angle $\Omega \equiv \{ \alpha, \beta, \gamma \}$ and $\hat{R} (\Omega) = {\rm e}^{-i \alpha \hat{I}_z} {\rm e}^{-i \beta \hat{I}_y} {\rm e}^{-i \gamma \hat{I}_z}$ is the rotational operator applied to the many-body wave function.
$I$ is the total angular momentum, $M$ and $K$ are angular momentum projections along the $z$-axis in the laboratory frame and intrinsic frame, respectively.
For nuclei with reflection asymmetric shapes, the parity projection should also be included to build states with definite parities~\cite{Ring1980_TNMBP}.
The parity projection operator writes
\begin{equation}
  \hat{P}^\pi
  =
  \frac{1}{2} (1 + \pi \hat{P}),
\end{equation}
where $\hat{P}$ is the spatial reflection operation and $\pi=\pm 1$ is the parity.

With $\hat{P}^I_{MK}$ and $\hat{P}^\pi$, we can project out states with good angular momentum $I$, $M$ and parity $\pi$ from a RHB mean field wave function $|\Phi\rangle$,
\begin{equation}
  \ket{\Psi_M^{I\pi}}
  =
  \sum_{K} f_K^{I\pi} \hat{P}^\pi \hat{P}_{MK}^I \ket{\Phi}
  =
  \sum_{K} f_K^{I\pi} \ket{IMK\pi},
  \label{eq:projected_wf}
\end{equation}
where $f^{I\pi}_K$ are some undetermined constants.
Taking Eq.~(\ref{eq:projected_wf}) as the trial wave function and solving the variational problem against the variables $f^{I\pi}_K$, we end up with the Hill-Wheeler-Griffin equation~\cite{Ring1980_TNMBP}
\begin{equation}
  \sum_{K'} [ \mathcal{H}_{KK'}^{I\pi} - E_{\alpha}^{I\pi} \mathcal{N}_{KK'}^{I\pi} ] f_{K'\alpha}^{I\pi}
  =
  0,
  \label{eq:HW_equation}
\end{equation}
where the Hamiltonian kernel $\mathcal{H}$ and the norm kernel $\mathcal{N}$ are defined as
\begin{subequations}
\begin{equation}
\begin{split}
  \mathcal{H}_{KK'}^{I\pi}
  &=
  \bra{\Phi} H P^\pi P_{KK'}^I \ket{\Phi} \\
  &=
  \frac{2I+1}{16\pi^2} \int \left[ D_{KK'}^{I^*}(\Omega) + \pi (-1)^I D_{K-K'}^{I}(\Omega) \right]
  \bra{\Phi} \hat{H} \hat{R} (\Omega) \ket{\Phi} {\rm d}\Omega,
  \end{split}
  \label{eq:Hker}
\end{equation}
\begin{equation}
\begin{split}
  \mathcal{N}_{KK'}^{I\pi}
  &=
  \bra{\Phi} P^\pi P_{KK'}^I \ket{\Phi} \\
  &=
  \frac{2I+1}{16\pi^2} \int \left[ D_{KK'}^{I^*}(\Omega) + \pi (-1)^I D_{K-K'}^{I}(\Omega) \right]
  \bra{\Phi} \hat{R} (\Omega) \ket{\Phi} {\rm d}\Omega.
\end{split}
   \label{eq:Nker}
\end{equation}
\end{subequations}
Note that the expressions of the kernels are derived by considering the $V_4$ symmetry. By discretizing the group space, the problem turns to the calculation of the norm overlap $n(\Omega)$ and the Hamiltonian overlap $h(\Omega)$ as functions of the Euler angle $\Omega$,
\begin{subequations}
  \begin{align}
    n (\Omega) &= \bra{\Phi}         \hat{R} (\Omega) \ket{\Phi}, \\
    h (\Omega) &= \bra{\Phi} \hat{H} \hat{R} (\Omega) \ket{\Phi} /n(\Omega).
  \end{align}
\end{subequations}
We adopt the method introduced in Ref. \cite{Yao2009_PRC79-044312} to calculate the overlaps in the ADHO basis using the generalized Wick's theorem.

We follow the standard way of solving the Hill-Wheeler-Griffin equation (\ref{eq:HW_equation}) as discussed in Ref. \cite{Ring1980_TNMBP}. The first step is to diagonalize the norm kernel
\begin{equation}
  \sum_{K'} \mathcal{N}_{KK'}^{I\pi} u_{m}^{IK'\pi}
  =
  n_m^{I\pi} u_{m}^{IK\pi}.
\end{equation}
The normalized vectors can be used to build orthonormalized basis
\begin{equation}
  \ket{m}
  =
  \frac{1}{\sqrt{n_m^{I\pi}}} \sum_K u_m^{IK\pi} \hat{P}^\pi \hat{P}_{MK}^I \ket{\Phi}.
\end{equation}
Eigen vectors with small eigen values $n^{I\pi}_m$ may induce numerical instabilities.
To solve this problem, we introduce a cutoff $\chi = 10^{-3}$ and only consider eigenstates with eigenvalues $n_m^{I\pi} > \chi$.
In the othornormalized basis $|m\rangle$ the Hill-Wheeler-Griffin equation becomes a standard eigenvalue problem,
\begin{equation}
  \sum_{m'} \braket{m | \hat{H} | m'} g_{m'}^{I\pi}
  =
  E^{I\pi} g_{m}^{I\pi}.
\end{equation}
The eigen energies $E_{\alpha}^{I\pi}$ can be solved by matrix diagonalization and the wave function components $f^{I\pi}_K$ can be obtained from the corresponding eigenvectors $g^{I\pi}_m$,
\begin{equation}
  f_K^{I\pi}
  =
  \sum_{m,n_m^{I\pi} \neq 0} \frac{g_m^{I\pi}}{\sqrt{n_m^{I\pi}}}
  u_m^{IK\pi}.
\end{equation}
With the projected wave functions, the reduced transition probabilities $B(E\lambda)$ from $J_i$ state to $J_f$ state writes
\begin{equation}
\begin{split}
  B(E \lambda;J_i \rightarrow J_f)
  &=
  \frac{1}{2J_i+1} \sum_{M_i M_f \mu}
  \big| \bra{J_f,M_f} \hat{Q}_{\lambda\mu} \ket{J_i,M_i} \big|^2 \\
  &=
  \frac{1}{2J_i+1} \big| \bra{J_f |} \hat{Q}_{\lambda} \ket{| J_i} \big|^2.
\end{split}
\end{equation}
The reduced matrix element has the following expression
\begin{equation}
  \bra{J_f |} \hat{Q}_{\lambda} \ket{| J_i}
  =
  \sum_{K_i K_f}
  f_{K_f}^{I_f \pi_f} f_{K_i}^{I_i \pi_i} \left[ Q_\lambda \right]_{K_f K_i}^{J_f J_i},
\end{equation}
where
\begin{equation}
   \left[ Q_\lambda \right]_{K_f K_i}^{J_f J_i}
   =
   \frac{(2J_i + 1)(2J_f + 1)}{8\pi^2} (-1)^{J_f - K_f} \sum_{\mu M}
   \left( \begin{array}{ccc}
     J_f & \lambda & J_i \\
    -K_f & \mu     & M \end{array} \right)
   Q_{\lambda \mu},
\end{equation}
in which
\begin{equation}
  Q_{\lambda \mu}
  =
  \int {\rm d} \Omega
  D_{M K_i}^{J_i *} (\Omega) \bra{\Phi} \hat{Q}_{\lambda \mu} \hat{R} (\Omega) \ket{\Phi}.
\end{equation}

\section{Results and Discussions}\label{sec:result}
In this section, we apply the p-MDCRHB model to study the low-lying excitations of the nucleus $^{12}$C, which is a typical nucleus with possible $\alpha$ clustering structures.
Among the proposed clustering states, the second $0^+$ state of $^{12}$C, namely the Hoyle state, is the most interesting because it plays a crucial role in the stellar nuclear synthesis \cite{Hoyle1954_AJSS1-121}.
The Hoyle state is a resonance consisting of three interacting $\alpha$ clusters lying above the three-$\alpha$ breakup threshold~\cite{Freer2014_PPNP78-1}.
The energy and width of this resonance play an essential role in synthesizing three $\alpha$'s into $^{12}$C in the main sequence stars.
The abundance of the carbon elements would be insufficient for supporting life in our universe if the Hoyle state has a different energy or does not exist at all.
Furthermore, several recent studies of the ground state band indicate that the 3-$\alpha$ structure may already appear in the ground state of $^{12}$C \cite{Bijker2016_PS91-073005,Kanada-Enyo2007_PTP117-655}.
A $5^-$ state was measured in the ground state band~\cite{Marin-Lambarri2014_PRL113-012502}, providing an evidence of a triangular $D_{3h}$ symmetry.


In MDCRHB model and its extensions, the system respects the mirror symmetries about the $x$-$z$ and $y$-$z$ planes.
One interesting case which these models specialize is the clustering structure in the light $\alpha$-conjugate nuclei.
Here the $\alpha$ clusters can form various molecular-like structures and the possible shapes include rod, triangle, tetrahedron, octahedron, etc.
Most of these shapes can not be described by conventional multipole expansion method that only includes either triaxial or octupole deformations.
Further, the structure might be different for ground state and excited state, a mean field description is too simplified and we must take into account the symmetry restorations with projection techniques.

\begin{figure}[htbp]
  \centering
  \includegraphics[width=8cm]{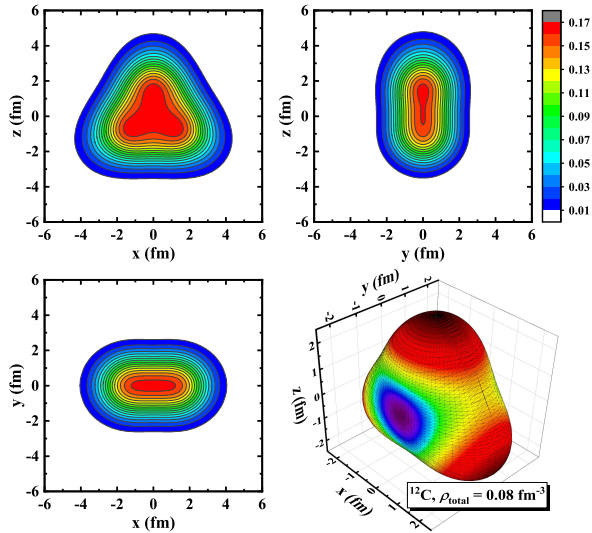}
  \caption{Density profiles of $^{12}$C constrained for $\lambda_3=0.8$.
  The contour plots display the total density $\rho_{\mathrm{total}}$ in $x$-$z$, $y$-$z$ and $x$-$y$ planes.
  The 3-D plot shows the isosurface of $\rho_{\mathrm{total}}$=0.08 fm$^{-3}$.}
  \label{Fig:C12_surface_contour}
\end{figure}

In this section, we apply the p-MDCRHB model to study the ground state and low-lying excitations of $^{12}$C nucleus with an intrinsic triangular shape.
To generate such a configuration, we define a triangular deformation constraint in the mean field calculations.
In analogy with the usual multipole deformation constraint Eq.~(\ref{eq:multipole_constraint}), we use the ansatz
\begin{equation}
  \lambda_{3}
  =
  \frac{4 \pi}{3 N R^3} \langle S_3 \rangle,
\end{equation}
where the ``triangular moment operator'' is defined as
\begin{equation}
S_3 = (z^2 + \rho^2 \cos^2 \varphi)^{3/2} \cos (3\theta),
\label{eq:tri_op}
\end{equation}
with $\rho$, $\varphi$ and $z$ the cylindrical coordinates and $\theta$ is the polar angle on the $x$-$z$ plane.
We can verify that Eq.~(\ref{eq:tri_op}) has the desired symmetry.
If rotated around $y$-axis, the operator $S_3$ has a period $2\pi/3$ and the maxima point to the directions along which the nucleus is stretched.
The parameter $\lambda_3$  is the counterpart of the multipole deformation parameters $\beta_{\lambda\mu}$.
When constrained to large $\lambda_3$, the operator $S_3$ elongates the nucleus in directions $\theta=0, 2\pi/3, 4\pi/3$ and compress it in other directions.
The resulting shape would be a regular triangle with three $^{4}$He at the vertices.
In MDCRHB model, the triangular constraint can be implemented in exactly the same way with the multipole moment constraint.
In Fig.~\ref{Fig:C12_surface_contour} we show the density profiles of $^{12}$C constrained to $\lambda_3=0.8$.
We present the contours in all three coordinate planes $x$=0, $y$=0, $z$=0 and a 3-D isosurface.
The densities are consistent with what we expect from the structure of the triangular moment constraint.
The energetically preferred directions are distributed evenly in the $x$-$z$ plane.
With much stronger constraining potential individual $\alpha$ clusters can form and be squeezed out of the nucleus.

\begin{figure}[htbp]
  \centering
  \includegraphics[width=8cm]{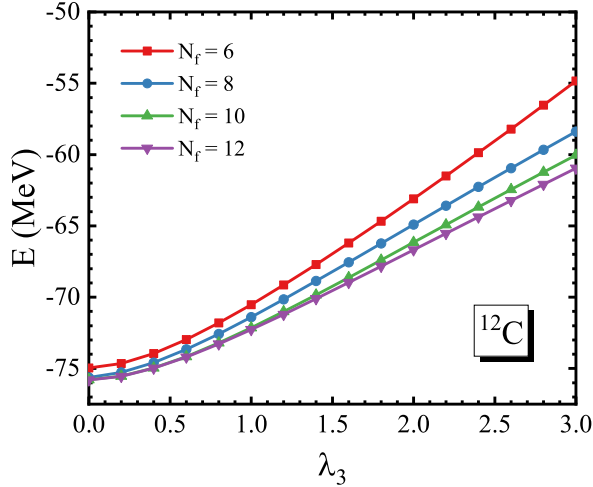}
  \caption{The energy of $^{12}$C as functions of the triangular shape parameter $\lambda_3$ calculated with different basis truncations in MDCRHB model.
  The squares, circles, up triangles and down triangles denote the results obtained with major shell truncation $N_f$=6, 8, 10 and 12, respectively.
  The center of mass correction to the binding energy is not included.}\label{Fig:conv_check_Nf}
\end{figure}

In mean field calculations, the Dirac equation is solved by expanding the wave functions in an ADHO basis with a major shell truncation $N_f$.
For Dirac spinors we use $N_f$ shells to expand the large component and $N_f + 1$ shells to expand the small component in order to remove the spurious states.
Firstly, we examine the convergence of the mean field results against the basis truncation $N_f$.
In Fig.~\ref{Fig:conv_check_Nf} we present the potential energy surfaces (PESs) of $^{12}$C calculated with the triangular constraint using different values of $N_f$.
We found that all energy minima locate at $\lambda_3 = 0$ corresponding to a spherical shape.
There is no explicit triangular correlation in the mean field level.
As $N_f$ increases, differences between adjacent curves become smaller, signifying a clear convergence pattern.
At the ground state $\lambda_3=0$, the calculated binding energy already converges at $N_f=8$, with the truncation error smaller than 0.1 MeV estimated by comparing to the result at $N_f=12$.
For larger $\lambda_3$, the mean field potential assumes a triangular shape, differing from that of the harmonic oscillator basis.
In this case we observe a slower convergence.
For example, the mean field energies for $N_f=10$ and $N_f=12$ differ by about 1 MeV for $\lambda_3 \approx 3.0$.
Nevertheless, the qualitative behaviours of the PES's are not changed when $N_f$ is large enough.
Concerning the big computational cost of the 3-D angular momentum projection, in this illustrative work we use $N_f=8$ in all following calculations.
For more realistic calculations we need to either employ a triangular harmonic oscillator basis or increase $N_f$ until all truncation errors are smaller than the desired precision.



\begin{figure}[htbp]
  \centering
  \includegraphics[width=16cm]{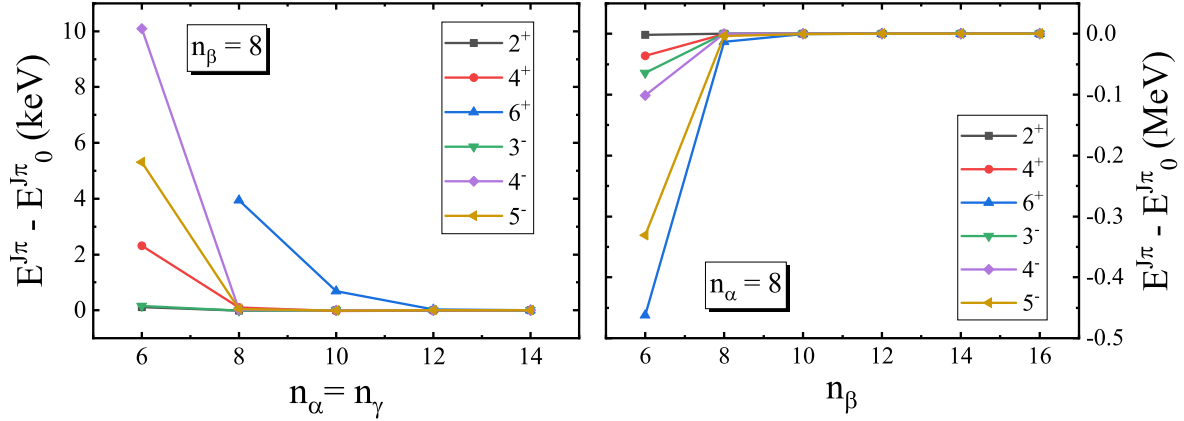}
  \caption{The projected energies of $^{12}$C calculated with different mesh size for SO(3) group integral.
  For each curve, the energies are relative to the corresponding converged values $E_0^J$ calculated at large mesh sizes.
  All calculations are performed with a mean field wave function constrained to $\lambda_3=1.2$.
  $n_\alpha$, $n_\beta$ and $n_\gamma$ are number of mesh points for integrating Euler angles $\alpha$, $\beta$ and $\gamma$, respectively.
  Squares, circles, up triangles, down triangles, diamonds and left triangles denote results projected to $J^\pi$=$2^+$, $4^+$, $6^+$, $3^-$, $4^-$, $5^-$, respectively.
  }
  \label{Fig:C12_conv_check_nmesh}
\end{figure}

For angular momentum and parity projections, the Hamiltonian kernel Eq.~(\ref{eq:Hker}) and the norm kernel Eq.~(\ref{eq:Nker}) are calculated by integrating over the SO(3) group space parameterized by the Euler angles $(\alpha,\beta,\gamma)$.
These integrals are performed numerically by the Gauss-Legendre integral with numbers of mesh points $(n_\alpha, n_\beta, n_\gamma)$ in $\alpha$, $\beta$ and $\gamma$ directions, respectively.
As most of the computational cost comes from the calculation of the Hamiltonian overlap and norm overlap for each Euler angle, it is desirable to reduce the total number of integral mesh points for a given precision.
Here we make a convergence check against the number of mesh points in the group space.
In Fig. \ref{Fig:C12_conv_check_nmesh}, we show the projected energies as functions of $n_\alpha$, $n_\beta$ and $n_\gamma$.
On the left panel, we set $n_\beta=8$ and study the projected energies for $6\leq n_\alpha=n_\gamma\leq14$, where $E^{J\pi}$ denotes the energy of the lowest $\ket{J^\pi}$ state with corresponding Euler angle mesh size, and $E^{J\pi}_0$ represents the energy with the largest mesh size.
we see that the influence of $n_\alpha$ and $n_\gamma$ on the projected energies for $J<6$ is less than 10 keV, which is negligible compared with the excitation energies of the order of MeV.
In contrast, the right panel shows that the number $n_\beta$ has a rather large influence on the projected energies.
All projected energies up to $6^+$ achieve relative precisions less than $10^{-6}$ when $n_\beta \geq 10$.
In this work, we choose the mesh size $n_\alpha = n_\gamma = 12$ and $n_\beta = 14$, with which the numerical integral in the group space can be evaluated rather precisely for $J\leq6$.

\begin{figure}[htbp]
  \centering
  \includegraphics[width=12cm]{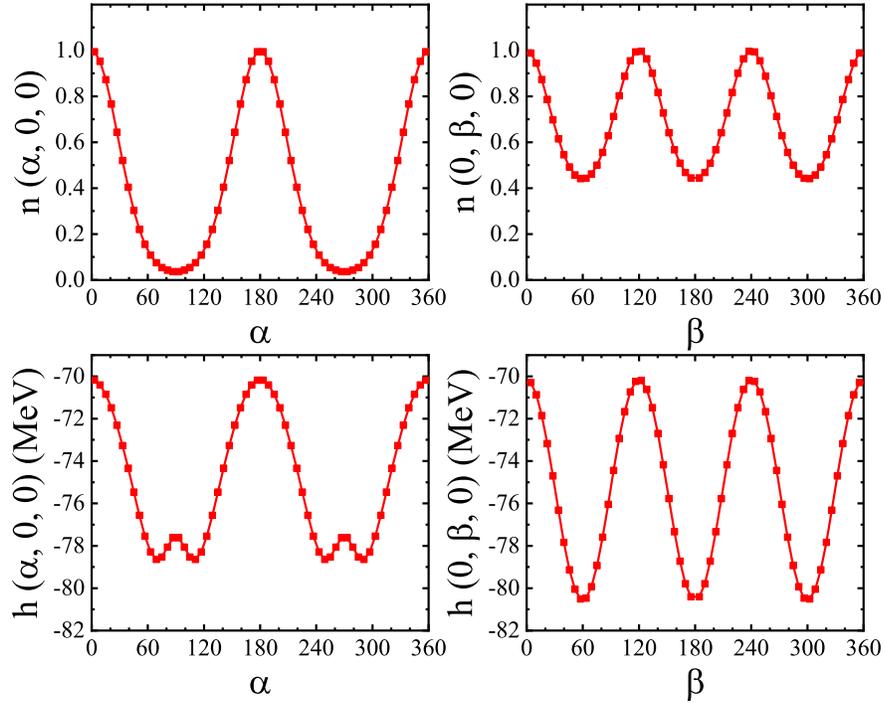}
  \caption{The norm and Hamiltonian overlaps of $^{12}$C as functions of the Euler angles $\alpha$ and $\beta$, calculated with a mean field wave function constrained to $\lambda_3 = 1.2$. $\beta=\gamma=0$ on the left panels and $\alpha = \gamma = 0$ on the right panels.}\label{Fig:C12_overlap}
\end{figure}

The constraint Eq.~(\ref{eq:tri_op}) produces a triangular shape with the 3-fold axis pointing along the $y$-axis and one of the 2-fold axes pointing along the $z$-axis.
For angular momentum projection we rotate the mean field wave function to Euler angles $(\alpha, \beta,\gamma)$ and calculate its overlap with the original wave function.
These norm overlaps should exhibit periodicity if the rotations are about the symmetry axes of the intrinsic shape, which can be examined numerically.
In Fig.~\ref{Fig:C12_overlap} we show the norm overlaps for different Euler angles.
All results are calculated with a mean field wave function constrained to $\lambda_3=1.2$.
In left panels we set $\beta=\gamma=0$ and rotate the nucleus around the $z$-axis by increasing $\alpha$ from $0^\circ$ to $360^\circ$.
As the $z$-axis is a 2-fold axis of the triangle, we found a period of $180^\circ$ for both the norm overlap and Hamiltonian overlap.
In right panels we set $\alpha=\gamma=0$, then variation of $\beta$ corresponds to a rotation around the $y$-axis.
In this case we observe a period of $120^\circ$ for both overlaps, which is consistent with the fact that $y$-axis is a 3-fold axis of the triangle.
In all subplots, the overlaps return to its original values at $\alpha,\beta,\gamma=0$ after rotating by the corresponding period.
This is the direct consequence of the invariance of the nuclear shape under these symmetry operations.

The spectrum of a triangular rotor can be mostly determined by the group theory analysis.
The triangle shape corresponds to a point group $D_{3h}$, which has 12 elements and 6 irreducible representations (\textit{irreps}) $A_1^\prime$, $A_2^\prime$, $E^\prime$, $A_1^{\prime\prime}$,
$A_2^{\prime\prime}$ and $E^{\prime\prime}$.
As $D_{3h}$ is a subgroup of the $O(3)$ group, we can decompose the \textit{irreps} of $O(3)$ group into \textit{irreps} of the $D_{3h}$ group~\cite{Hamermesh1962_GroupTheory,Gozdz1996_APPB27-469,Tagami2013_PRC87-054306}.
The $O(3)$ group has infinitely many \textit{irreps} corresponding to different angular momentum $I$ and parity $\pi$.
In Table~\ref{tab:n_states_D3h} we present the decomposition of the $O(3)$ \textit{irreps} into the $D_{3h}$ \textit{irreps}.
In each column we show the number of each $D_{3h}$ \textit{irrep} appearing in the decomposition.
For example, the identity representation corresponds to $0^+$ for $O(3)$ group and is marked by $A_1^\prime$ for $D_{3h}$ group, while the \textit{irrep} of $O(3)$ with $I^\pi=1^+$ can be written as a direct sum of $A_2^\prime$ and $E^{\prime\prime}$ representations.

For any state of the triangular rotor with quantum numbers $I^\pi$, the wave function transforms under the corresponding \textit{irrep} of the $O(3)$ group.
If we rotate the rotor with an element of the $D_{3h}$ group in the intrinsic frame, nothing is actually changed, thus the wave functions should be scalars for these rotations.
In the language of the group theory, the allowed states all belong to the $A_1^\prime$ \textit{irrep} of the intrinsic $D_{3h}$ group.
If an angular momentum and parity combination $I^\pi$ does not contain any $A_1^\prime$ \textit{irrep}, it should not appear in the spectrum.
In other words, the $I^\pi$ states with non-zero entries in the first row of Table~\ref{tab:n_states_D3h} are all states allowed by the symmetry analysis.
In summary, the lowest energy levels of a triangular rigid rotor should be $I^\pi=0^+,2^+,3^-,4^+,4^-,5^-,6^+,6^-,...$.
Other angular momentum and parity combinations should be consequences of the intrinsic excitations.

\begin{table}[htbp]
\centering
\caption{The decomposition of the \textit{irreps} of the $O(3)$ group with quantum numbers $I^\pi$ into \textit{irreps} of the $D_{3h}$ group.
In each column the numbers are the degeneracies.
}\label{tab:n_states_D3h}
\begin{tabular}{ccccccccccccccccc}
\hline
\hline
   $I^\pi$  & $0^+$ & $1^+$ & $2^+$ & $3^+$ & $4^+$ & $5^+$ & $6^+$ & $7^+$ & $0^-$ & $1^-$ & $2^-$ & $3^-$ & $4^-$ & $5^-$ & $6^-$ & $7^-$  \\
\hline
   $A'_1$   &     1&     0&     1&     0&     1&     0&     2&     1&     0&     0&     0&     1&     1&     1&     1&     1  \\
   $A'_2$   &     0&     1&     0&     1&     0&     1&     1&     2&     0&     0&     0&     1&     1&     1&     1&     1  \\
   $E'$     &     0&     0&     1&     1&     2&     2&     2&     2&     0&     1&     1&     1&     1&     2&     2&     3  \\
   $A''_1$  &     0&     0&     0&     1&     1&     1&     1&     1&     1&     0&     1&     0&     1&     0&     2&     1  \\
   $A''_2$  &     0&     0&     0&     1&     1&     1&     1&     1&     0&     1&     0&     1&     0&     1&     1&     2  \\
   $E''$    &     0&     1&     1&     1&     1&     2&     2&     3&     0&     0&     1&     1&     2&     2&     2&     2  \\
\hline
\hline
\end{tabular}
\end{table}

On the one hand, the group theory analysis is strict and can be generalized to any other intrinsic shapes.
For example, we can predict the low-lying states of a tetrahedral rotor and compare the results with that of the $^{16}$O nucleus, which was predicted to exhibit a four-$\alpha$ cluster structure.
On the other hand, there are also less abstract ways of understanding the essentiality of the nuclear rotational spectrum.
For a triangular rotor, a stable rotation can only occur around its symmetry axes.
The wave function for angular momentum $I$ is $\psi(\varphi)=\exp(iI\varphi)$ with $\varphi$ the rotational angle.
For a $n$-fold axis we must have $\psi(\varphi + 2\pi/n) = \psi(\varphi)$, thus among all integer values of $I$, only integer multiples of $n$ are allowed.
For example, the rotations about a 2-fold axis generate angular momenta $I=0, 2, 4, 6,...$, while the rotations about a 3-fold axis give the angular momenta $I=0, 3, 6,...$.
Further, the parity of the rotational wave function can be determined by $\pi=(-1)^I$.
The total rotational energy of a triangular rotor is the sum of that around the 2-fold and 3-fold axes.
If there is no rotation around the 3-fold axis, we have the spectrum $I^\pi=0^+, 2^+, 4^+, 6^+,...$ If there is one unit of quantized angular momentum around the third axis, we have states $I^\pi=3^-,4^-,5^-,6^-,...$.
Here the states $4^-$ and $5^-$ can be understood as a rotation around the axis in between the 2-fold axis and the 3-fold axis, which is already a complex motion in the classical level.



\begin{figure}[htbp]
  \centering
  \includegraphics[width=10cm]{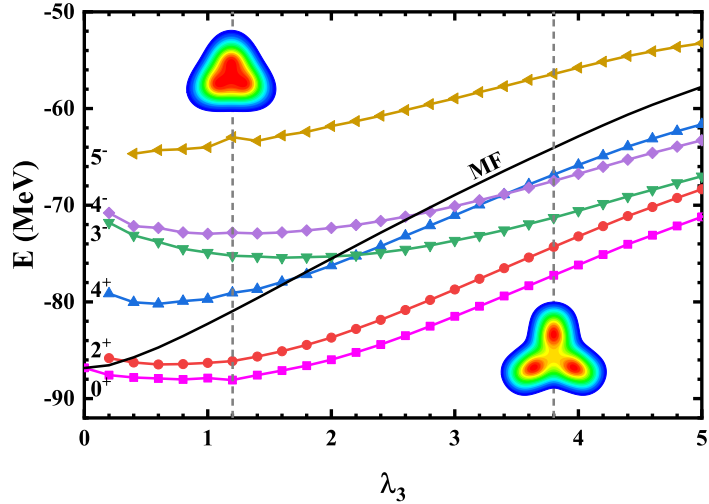}
  \caption{
  The mean field and the projected potential energies of $^{12}$C versus the triangular deformation $\lambda_3$.
  The squares, circles, down triangles, up triangles, diamonds and left triangles represent $I^\pi=0^+$, $2^+$, $3^-$, $4^+$, $4^-$ and $5^-$, respectively.
  The mean field energies are displayed as black solid line.
  The insets show the mean field density profiles at $\lambda_3 = 1.2$ and $3.8$.}\label{Fig:C12_PES_spectra}
\end{figure}

Next we turn to the projected potential energy surfaces (PES's).
We first generate mean field configurations with $\lambda_3$ varying from 0.0 to 5.0.
The results describe the disintegration of $^{12}$C into three $\alpha$ clusters.
In Fig.~\ref{Fig:C12_PES_spectra} we show the projected energies as well as mean field energies for all $\lambda_3$ in this interval with a step size $\Delta \lambda_3=0.2$.
Here we only show the lowest energy levels with $I^\pi=0^+$, $2^+$, $3^-$, $4^\pm$ and $5^-$.
The mean field PES gives a ground state at the spherical shape $\lambda_3=0$.
The energy increases quickly as three $\alpha$'s are separated.
For $\lambda_3=1.0$ the deformation energy is more than 5 MeV, preventing the dissociation of the nucleus.
By projecting out the spurious rotational energies, we see that the PES for $0^+$ state is quite different from the mean field PES.
The $0^+$ PES is lower than the mean field PES and becomes very flat for $\lambda_3 < 2.0$.
We find a minimum at $\lambda_3=1.2$, which is very shallow that the energy at $\lambda_3=0$ is only about 1 MeV higher.
In the insets of Fig.~\ref{Fig:C12_PES_spectra} we show the density profiles from the mean field calculations at $\lambda_3=1.2$ and $\lambda_3=3.8$.
The first inset corresponds to the ground state identified with the $0^+$ PES.
Although at this point there is no emerging clustering structure, we can see a clear triangular shape from the total densities.
At $\lambda_3=3.8$ the deformation is so strong that the clustering structure is very pronounced.
Theoretically for extremely large $\lambda_3$ the $^{12}$C nucleus would break up into three $\alpha$ particles and the energy should converge to three times the $^{4}$He binding energy.
However, for such light nuclei we might need a very large center of mass correction and the mean field model might not be a good approximation.
Even though there are still a lot of difficulties, the extremely large $\lambda_3$ deformation still deserves some further investigations.
This three-$\alpha$ disintegration can be viewed as an inverse reaction of the famous $3\alpha\rightarrow^{12}$C reaction and may be related to the Hoyle state.
Whether the mean field model is adequate for describing such a process and what is lack is thus an interesting topic.
Although in recent years the \textit{ab initio} calculations are more popular in studying the clustering structures, mean field models have a clear physical pictures and have proven successful on this topic.
The p-MDCRHB model we presented here can describe various $\alpha$ cluster configurations and can serve as a starting point for beyond-mean-field calculations in this direction.

The PES's for higher angular momenta tell us how the $\alpha$ clusters move about the rotational axes.
At $\lambda_3=0$ the system is spherical and there is no rotation.
For larger $\lambda_3$, the $2^+$ and $4^+$ states describe rotations around the 2-fold axes, the corresponding
PES's have mimina at relatively small $\lambda_3$.
The $3^-$ and $4^-$ states contain rotations around the 3-fold axis, thus the centrifugal force prefers a larger $\lambda_3$ which also means a larger moment of inertia.





\begin{table}[htbp]
\centering
\caption{Reduced $E2$ transition probabilities from states $I^\pi_i$ to states $I^\pi_f$ and the corresponding excitation energies of $^{12}$C.
The calculation is made for $\lambda_3$=1.2 corresponding to the energy minimum of the $0^+$ PES shown in Fig. 5.
The experimental $B(E2)$ value is taken from Ref.~\cite{Ajzenberg-Selove1990_NPA506-1}.}\label{tab:BE2}
\begin{tabular}{ccccccc}
\hline
\hline
   $I^\pi_i$ & $E_{xi}^\mathrm{th}$ (MeV) & $E_{xi}^\mathrm{expt}$ (MeV) & $I^\pi_f$ & $E_{xf}^\mathrm{th}$ (MeV) & $B(E2)^\mathrm{th}$ (e$^2$ fm$^4$) & $B(E2)^\mathrm{expt}$ (e$^2$ fm$^4$) \\
\hline
   $2^+$   & 1.951  &  4.440 & $0^+$ &  0.000 &  9.72 & 7.6 $\pm$ 0.4 \cite{Ajzenberg-Selove1990_NPA506-1} \\
\hline
   $4^+$   & 9.015  & 14.079 & $2^+$ &  1.951 & 16.28 & \\
\hline
   $4^-$   & 15.253 && $3^-$ & 12.864 & 20.95 & \\
\hline
   $5^-$   & 19.915 && $3^-$ & 12.864 &  6.78 & \\
           &        && $4^-$ & 15.253 & 18.90 & \\
\hline
   $6^+_1$ & 26.734 && $4^+$ &  9.015 & 26.04 & \\
\hline
   $6^-$   & 25.131 && $4^-$ & 15.253 & 13.38 & \\
           &        && $5^-$ & 19.915 & 15.01 & \\
\hline
\hline
\end{tabular}
\end{table}


\begin{table}[htbp]
\centering
\caption{Reduced $E3$ transition probabilities from states $I^\pi_i$ to states $I^\pi_f$ and the corresponding excitation energies of $^{12}$C.
The calculation is made for $\lambda_3$=1.2 corresponding to the energy minimum of the $0^+$ PES shown in Fig. 5.
The experimental $B(E3)$ value is taken from Ref. \cite{Ajzenberg-Selove1990_NPA506-1}.}\label{tab:BE3}
\begin{tabular}{ccccccc}
\hline
\hline
   $I^\pi_i$ & $E_{xi}^{\mathrm{Th}}$ (MeV) & $E_{xi}^\mathrm{Expt}$ (MeV) & $I^\pi_f$ & $E_{xf}^{\mathrm{Th}}$ (MeV) & $B(E3)^\mathrm{Th}$ (e$^3$ fm$^6$) & $B(E3)^\mathrm{Expt}$ (e$^3$ fm$^6$) \\
\hline
   $3^-$   & 12.864 &  9.641 & $0^+$ &  0.000 & 145.81 & 103 $\pm$ 17 \cite{Ajzenberg-Selove1990_NPA506-1} \\
           &        &        & $2^+$ &  1.951 & 260.25 & \\
           &        &        & $4^+$ &  9.015 &  53.53 & \\
\hline
   $4^-$   & 15.253 & 13.35 & $2^+$ &  1.951 & 268.69 & \\
           &        &        & $4^+$ &  9.015 & 198.55 & \\
\hline
   $5^-$   & 19.915 & 22.4(2) & $2^+$ &  1.951 &  96.28 & \\
           &        && $4^+$ &  9.015 & 318.35 & \\
\hline
   $6^+_1$ & 26.734 && $4^-$ & 15.253 &  33.95 & \\
           &        && $5^-$ & 19.915 & 129.86 & \\
           &        && $6^-$ & 25.131 & 250.19 & \\
\hline
   $6^+_2$ & 27.234 && $3^-$ & 12.864 & 480.28 & \\
           &        && $4^-$ & 15.253 & 219.26 & \\
           &        && $5^-$ & 19.915 &  90.68 & \\
           &        && $6^-$ & 25.131 &  30.99 & \\
\hline
   $6^-$   & 25.131 && $4^+$ &  9.015 & 274.90 & \\
\hline
\hline
\end{tabular}
\end{table}

The energy spectra of $^{12}$C at large $\lambda_3$ can be understood by comparing them to the spectra of a triangular rigid rotor discussed above.
Taking such a rotor with equal masses on the three vertices, we can compute the moment of inertia $\mathcal{M}$ along every principle axis.
For the orientation shown in Fig.~\ref{Fig:C12_surface_contour}, the principle axes are simply the three coordinate axes $x$, $y$ and $z$.
It is straightforward to prove that 2$\mathcal{M}_x$ = 2$\mathcal{M}_z$ = $\mathcal{M}_y$ = $\mathcal{M}$.
Only considering rotations about the center of mass, we can write the Hamiltonian as
\begin{equation}
  \hat{H} = \frac{\hat{I}_x^2}{\mathcal{M}} + \frac{\hat{I}_y^2}{2 \mathcal{M}}
  + \frac{\hat{I}_z^2}{\mathcal{M}}
  = \frac{\hat{I}^2}{\mathcal{M}} - \frac{\hat{I}_y^2}{2 \mathcal{M}},
  \label{eq:rotor_Hamiltonian}
\end{equation}
with $\hat{I}_x$, $\hat{I}_y$ and $\hat{I}_z$  the angular momentum components in the intrinsic frame and $\hat{I}^2$ the total angular momentum.
As $\hat{H}$, $\hat{I}^2$ and $\hat{I}_y^2$ commute with each other, we can take the total energy $E$, total angular momentum $I$ and the angular momentum component $K$ as a group of good quantum numbers specifying the eigenstates.
Here $I(I+1)$ is the eigenvalue of $\hat{I}^2$ and $K$ is the eigenvalue of $\hat{I}_y$.
Rotation around the intrinsic $y$-axis by $2\pi/3$ takes the nucleus back to its original position while produces a phase $\exp(i2\pi K/3)$.
This phase should be unit one for any physically allowed state, thus we have $K=3n$ with $n$ an integer.
For a state $|IK\alpha\rangle$ with $\alpha$ quantum numbers other than $I$ and $K$, we can calculate the energy according to Eq.~(\ref{eq:rotor_Hamiltonian}),
\begin{equation}
  E = \langle IK\alpha| \hat{H} |IK\alpha \rangle
    = \frac{1}{\mathcal{M}} [I(I+1) - K^2/2].
    \label{eq:Energy_Spectrum}
\end{equation}
The spatial reflection $P$ combined with a rotation around the 3-fold axis by $180^\circ$ reset the system.
The operator $P$ produce a factor of parity $\pi$ and the rotation gives an additional phase of $(-1)^K$, thus we have $\pi=(-1)^K$ for all physically allowed states.

The energy spectrum of Eq.~(\ref{eq:Energy_Spectrum}) can be classified according to the quantum number $K$.
For $K=0$ we have the ground state band $0^+$, $2^+$, $4^+$, ..., with positive parity.
For $K=3$ we have a negative parity band $3^-$, $4^-$, $5^-$, ...
The excitation energies relative to the $0^+$ state are
\begin{equation}
     E_x(4^+):E_x(4^-):E_x(3^-):E_x(2^+)
  =  3.333:2.583:1.25:1
\end{equation}
up to an overall multiplicative factor.
One important feature of this spectrum is that the $4^+$ energy is higher than the $4^-$ energy.
The reason is that the $4^+$ state is a pure rotation around the 2-fold axes, while $4^-$ state contains rotations around the 3-fold axis.
The moment of inertia for the latter rotation is relatively larger.
However, in Fig.~\ref{Fig:C12_PES_spectra} we see that the $4^+$ state is much lower than $3^-$ and $4^-$ for small $\lambda_3$, only for extremely large $\lambda_3$ the projected energies can be explained by the rigid rotor model.
The reason is twofold.
First, the relation $\mathcal{M}_y$ = 2$\mathcal{M}_z$ used in deriving the rigid rotor spectrum is only approximately satisfied for large deformations where the masses of $\alpha$ clusters are concentrated at the vertices.
For small $\lambda_3$ we have a triangular density distribution without clustering and the moments of inertia along different axis are not so different.
Second, the angular momentum and parity projections capture the full quantum mechanical elements.
The impact of the intrinsic states are reflected by the calculated norm kernel and Hamiltonian kernel.
Only for large $\lambda_3$ we can omit the intrinsic structures and use the Gaussian overlap approximation (GOA) to treat the $\alpha$ clusters as point particles.

In Table~\ref{tab:BE2} and Table~\ref{tab:BE3} we compare the low-lying spectrum and reduced transition probabilities $B(E2)$ and $B(E3)$ of $^{12}$C with the experiments.
The results are calculated by projecting a mean field state constrained to $\lambda_3=1.2$, which corresponds to the minimum of the $0^+$ PES in Fig~\ref{Fig:C12_PES_spectra}.
The experimental values are well reproduced at this level of approximations.
These results can be improved from various aspects.
First, as the PES's are soft at small $\lambda_3$, in Fig.~\ref{Fig:C12_PES_spectra} we see that the $\lambda_3$ value minimizing the projected energy has a strong dependence on $I^\pi$, which should be considered in calculating the energy spectrum.
Second, the soft PES's suggest we introduce the generating coordinate method (GCM) to mix different deformations.
Lastly, other configurations like the rod shape might also play an important role in the low-lying excitations and should be considered in building the low-lying states with good quantum numbers.

\begin{figure}[htbp]
  \centering
  \includegraphics[width=10cm]{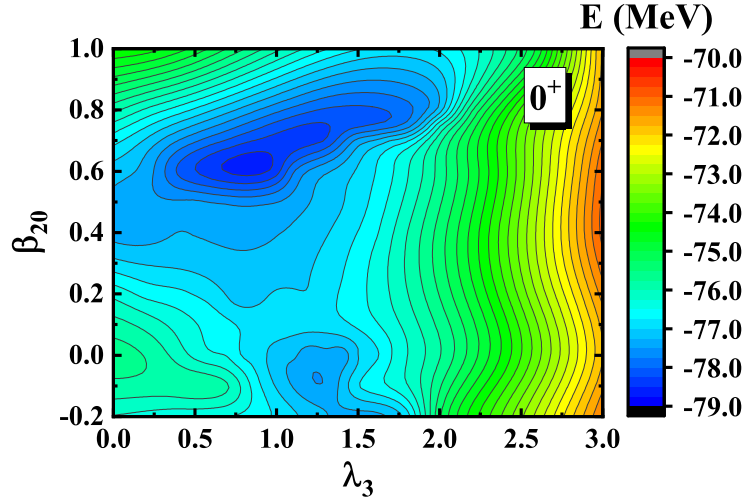}
  \caption{The 2-D potential energy surface of $^{12}$C projected to $0^+$ state.
  The $\lambda_3$ and $\beta_{20}$ constraints are imposed with $0 < \lambda_3 < 3.0,~\Delta \lambda_3 = 0.2$ and $-0.2 < \beta_{20} < 1.0,~\Delta \beta_{20} = 0.1$.}\label{Fig:C12_2D_PES}
\end{figure}

In Ref. \cite{Bijker2002_AP298-334}, the $\alpha$ arrangement in $^{12}$C is assumed to be an equilateral triangle.
In Ref.~\cite{Moriya2021_FBS62-46}, the author performed calculations with the orthogonality condition model (OCM) and the shallow potential model, and obtained different configurations for each model.
Due to insufficient experimental data, whether the intrinsic configuration of the ground state of $^{12}$C is a regular triangle remains an open questions~\cite{Moriya2021_FBS62-46}.
It is necessary to explore this point by performing multi-dimensional constrained calculations that include all important shape configurations.
In Fig.~\ref{Fig:C12_2D_PES} we show the projected $0^+$ PES of $^{12}$C with respect to the deformation parameters $\beta_{20}$ and $\lambda_3$.
Here we constrain $\lambda_3$ from 0.0 to 3.0 with a step size $\Delta \lambda_3 = 0.2$, constrain $\beta_{20}$ from $-0.2$ to $1.0$ with a step size $\Delta \beta_{20} = 0.1$.
The projected energies are much lower than the mean field energies for non-zero deformations.
The rotational energy correction defined as the difference between the $0^+$ energy and the mean field energy plays a key role in determining the ground state configurations.
On the $0^+$ PES, we found two energy minima.
The lowest one locates at $\lambda_3 = 0.8$ and $\beta_{20} = 0.6$, corresponding to a isosceles acute triangular shape,
while the shape isomer at $\lambda_3=1.2$ and $\beta_{20}=-0.1$ corresponds to an isosceles obtuse triangular shape.

\section{\label{sec:summary}Summary}
We have developed a projected multi-dimensionally constrained relativistic Hartree-Bogoliubov (p-MDCRHB) model.
In this model, the shape of the system is only restricted as mirror reflection symmetric about $x$-$z$ and $y$-$z$ planes.
In the mean field calculation, the RHB equation is solved in the axially deformed harmonic oscillator (ADHO) basis.
The parity and angular momentum projection is performed to restore the intrinsic symmetries.

This model is well suited for exploring exotic shapes like triangle and tetrahedron found in both the nuclear ground state and low-lying excited states.
One important scenario where these shapes play essential roles is the nuclear clustering.
The most important stellar nuclear reactions such as the production of carbon and oxygen depend heavily on the clustering structures of the reactant.
In this work we benchmark the pMDCRHB model by studying the clustering states in $^{12}$C.
We define a triangular moment constraint and study the corresponding triangular deformation $\lambda_3$.
This constraint generates a clustering state with three $\alpha$ clusters arranged on the vertices of a regular triangle.
We found that with the growth of $\lambda_3$, the spectra approach that of a regular triangular rigid rotor.
This was explained by the symmetry analysis and the rigid rotor Hamiltonian.
We also performed the 2D constrained calculations of $^{12}$C.
Two minima are found on the projected $0^+$ PES.
The softness of the PES's indicates that the configuration mixing GCM calculation is needed for a deeper understanding of the structure of $^{12}$C and the related nuclear reactions occurring in the stellar environment.

\section{\label{sec:Acknowledgments}Acknowledgments}
Helpful discussions with Xiang-Xiang Sun and Yu-Ting Rong are gratefully acknowledged. We thank Xiang-Xiang Sun and Shan-Gui Zhou for reading the manuscript and valuable suggestions. This work has been supported by NSAF (Grant No. U1930403), the National Key R\&D Program of China (Grant No. 2018YFA0404402), the National Natural Science Foundation of China (Grants No. 11525524, No. 12070131001, No. 12047503, and No. 11961141004), the Key Research Program of Frontier Sciences of Chinese Academy of Sciences (Grant No. QYZDB-SSWSYS013) and the Strategic Priority Research Program of Chinese Academy of Sciences (Grants No. XDB34010000 and No. XDPB15). The results described in this paper are obtained on the High-performance Computing Cluster of ITP-CAS and the ScGrid of the Supercomputing Center, Computer Network Information Center of Chinese Academy of Sciences.

\section*{Appendix: Calculation of the Rotation Matrix}

Since the rotation matrix is naturally written as Wigner $D$-function with spherical harmonic oscillators, the calculation of the rotation matrix can be simplified by expanding the ADHO basis with the spherical harmonic oscillator (SHO) basis.

An ADHO basis $\ket{\alpha}$ with quantum numbers $\{n_z,n_\rho,m_l,m_s\}$ reads
\begin{equation}
  \ket{\alpha}
  =
  \ket{n_z n_\rho m_l m_s}
  =
  C_L \phi_{n_z} (z) R_{m_l}^{n_\rho} (\rho,\varphi) \chi_{m_s},
\end{equation}
with
\begin{subequations}\label{Eqs:ADHO}
\begin{align}
  \phi_{n_z} (z)
  &=
  \frac{1}{\sqrt{b_z}} \frac{1}{\pi^{1/4} \sqrt{2^{n_z} n_z !}} H_{n_z} \left(\frac{z}{b_z}\right) {\rm e}^{-\frac{z^2}{2b_z}}, \\
  R_{m_l}^{n_\rho} (\rho,\varphi)
  &=
  \frac{1}{b_\rho} \sqrt{ \frac{2 n_\rho !}{(n_\rho + |m|)!} } \frac{1}{\sqrt{2\pi}} {\rm e}^{-\rho^2/(2b_\rho^2)}
  \left( \frac{\rho}{b_\rho} \right)^{|m|} L_{n_\rho}^{|m|} \left( \frac{\rho^2}{b_\rho^2} \right) {\rm e}^{im\varphi}.
\end{align}
\end{subequations}
$b_z$ and $b_\rho$ are the harmonic oscillator lengths along $z$ and $\rho$ directions which are defined as $b_z = 1/\sqrt{M\omega_z}$ and $b_\rho = 1/\sqrt{M\omega_\rho}$.

The SHO basis function reads
\begin{equation}
  \ket{n_r l j m}
  =
  \sum_{m_s} C_{m_l m_s j m}^{l \frac{1}{2}} \phi_{lm_l}^{n_r} (r,\Omega) \chi_{m_s},
\end{equation}
with $C_{m_l m_s j m}^{l \frac{1}{2}}$ the Clebsch-Gordon coefficient and
\begin{equation}\label{Eq:SHO}
  \phi_{lm_l}^{n_r} (r,\Omega)
  =
  \frac{1}{b_r^{3/2}} \sqrt{ \frac{2n_r!}{\Gamma (n_r+l+3/2)} } \left( \frac{r}{b_r} \right)^l
  \exp \left( -\frac{r^2}{2b_r^2} \right) L_{n_r}^{l+1/2} \left( \frac{r^2}{b_r^2} \right)
  Y_{lm} (\Omega).
\end{equation}
The transformation matrix can be written as
\begin{equation}
\begin{split}
  \braket{n_z n_\rho m_l m_s | n_r l j m_j }
  &=
  C_L^* C_{m_l m_s j m_j}^{l\frac{1}{2}} \delta_{m_l+m_s,m_j}
  \braket{\phi_{n_z} R_{m_l}^{n_\rho} | \phi_{l m_l}^{n_r}}\\
  &=
  C_L^* C_{m_l m_s j m_j}^{l\frac{1}{2}} \delta_{m_l+m_s,m_j}
  \braket{ n_z n_\rho m_l | n_r l m_l }.
\end{split}
\end{equation}
From Eq.~(\ref{Eqs:ADHO}) and Eq.~(\ref{Eq:SHO}), one can calculate the transformation matrix element, which has also been given in Ref.~\cite{Talman1970_NPA141-273}
\begin{equation}
  \braket{n_r l m | n_z n_\rho m}
  =
  \sum_\lambda
  \frac{ n_\rho ! (2l+1) 2^{n_\rho-n_r-\lambda} (-1)^\lambda (n_r+2\lambda) !}
  {\lambda! (n_\rho-\lambda)! (\lambda-n_\rho+n_r)! (2n_r+2\lambda-2n_\rho+2l+1)!!}.
\end{equation}
Thus, one obtains the relation between ADHO and SHO bases,
\begin{equation}
  \ket{\alpha}
  =
  \sum_{n_r l j m} C_{n_r l j m}^\alpha \ket{ n_r l j m }.
\end{equation}

The time reversal states of the spherical harmonic oscillator basis are obtained by applying the time reversal operator $\hat{T}$
\begin{equation}
  \hat{T} \ket{n_r l j m_j}
  =
  (-1)^{l+m_j+j} \ket{n_rlj\bar{m}_j}.
\end{equation}

The rotation matrix in the ADHO basis is calculated as
\begin{subequations}\label{Eq;Rmatrix}
\begin{align}
  \bra{\beta} \hat{R} (\Omega) \ket{\alpha}
  &=
  \sum_{n_r l j} \sum_{mm'} C_{n_r l j m}^{\beta *} D_{mm'}^j (\Omega) C_{n_r l j m'}^\alpha, \\
  \bra{\bar{\beta}} \hat{R} (\Omega) \ket{\alpha}
  &=
  \sum_{n_r l j} \sum_{mm'} (-1)^{l+m-j} C_{n_r l j m}^{\beta} D_{mm'}^j (\Omega) C_{n_r l j m'}^\alpha
\end{align}
\end{subequations}
By applying the properties of Wigner $D$-function, one finds the relation
\begin{equation}
  R_{\bar{\beta}\bar{\alpha}} = R_{\beta\alpha}^*, \quad
  R_{\beta\bar{\alpha}} = - R_{\bar{\beta}\alpha}^*, \quad
  \mathrm{for}~\alpha,\beta > 0,
  \label{eq:Matrix_symm}
\end{equation}
that means only half of $R$ matrix elements are needed in Eq.~(\ref{Eq;Rmatrix}), while the other half can be calculated directly from the relation (\ref{eq:Matrix_symm}).

%
%
%
%
%
%
%
%
%
%
%
%
%
%
%
%
%

\vspace*{2mm}

\bibliographystyle{CommTP}
\end{document}